\documentclass[review]{elsarticle}

\usepackage{lineno,hyperref}
\modulolinenumbers[1]
\usepackage{multirow}
\usepackage[utf8]{inputenc}
\usepackage{soul}
\usepackage{xcolor}
\usepackage{comment}
\usepackage{ulem}
\usepackage{makecell}

\journal{Physics Letters B}

\newcommand{\pp}{$p$+$p$}

\newcommand{\pT}{\ensuremath{p_\mathrm{T}}~}

\begin{document}

\begin{frontmatter}

\title{Measurements of the $Z^{0}/\gamma^{*}$ cross section and transverse single spin asymmetry in 510 GeV $p$$+$$p$ collisions}

\author{The STAR Collaboration\corref{mycorrespondingauthor}}

\begin{abstract}
The differential cross section for $Z^{0}$ production, measured as a function of the boson's transverse momentum ($p_{\mathrm{T}}$), provides important constraints on the evolution of the transverse momentum dependent parton distribution functions (TMDs). The transverse single spin asymmetry (TSSA) of the $Z^{0}$ is sensitive to one of the polarized TMDs, the Sivers function, which is predicted to have the opposite sign in $p+p$ $\rightarrow W/Z+ X$ from that which enters in semi-inclusive deep inelastic scattering.
In this Letter, the STAR Collaboration reports the first measurement of the $Z^{0}/\gamma^{*}$ differential cross section as a function of its \pT in $p$$+$$p$ collisions at a center-of-mass energy of 510 GeV, together with the $Z^{0}/\gamma^{*}$ total cross section. We also report the measurement of $Z^{0}/\gamma^{*}$ TSSA in transversely polarized $p$$+$$p$ collisions at 510 GeV.
\end{abstract}

\begin{keyword} $Z^{0}$, cross section, transverse momentum dependent parton distribution functions (TMDs), transverse single spin asymmetry (TSSA, $A_{N}$), Sivers function
\end{keyword}

\end{frontmatter}
\newpage
\section{Introduction}\label{sec:intro}
The internal structure of hadrons, described by their parton distribution functions (PDFs) \cite{Soper:1996sn}, is an important topic in theoretical, phenomenological, and experimental studies in nuclear physics. During the past decades, numerous efforts have been made to understand transverse momentum dependent parton distribution functions (TMDs)~\cite{Aybat:2011zv} which encode both the parton's longitudinal momentum fraction ($x$) and its intrinsic transverse momentum ($k_{\mathrm{T}}$). TMDs depict the density of partons in three dimensions \cite{PhysRevD.41.83,Collins:1992kk}, providing more detailed information on hadron structure than the one-dimensional collinear PDFs. There are eight leading-twist TMDs that are allowed by parity invariance \cite{Meissner:2009ww} of Quantum Chromodynamics (QCD). They cannot be calculated from first principles as it requires QCD calculations in a nonperturbative regime. Many observables in hard scattering experiments involving hadrons are related to TMDs. Utilizing factorization theorems, TMDs can be extracted through global fits of the cross section and other observables.

Observables related to TMDs require the measured transverse momentum component to be much smaller than the hard scale of the process. In semi-inclusive deep inelastic scattering (SIDIS), the hard scale is characterized by the square of the 4-momentum of the exchanged virtual photon ($Q^{2}$ = $-q^{2}$). If the measured transverse momentum of the outgoing hadron is small, $p^{h}_{\mathrm{T}}\ll Q$, then TMD factorization can be applied. TMDs can also be extracted from di-lepton production in Drell-Yan (DY) events \cite{Drell:1970wh} if the transverse momentum of the lepton pair is sufficiently small, $q_{\mathrm{T}}\ll Q$. In the $Z^{0}$ production events, $Q^{2}$ is the square of the $Z^{0}$ boson mass. On one hand, measuring the differential cross section as a function of transverse momentum for different processes tests the universality of TMDs and provides opportunities to study their $Q^{2}$ evolution.
Measurements of $p+p \rightarrow Z^{0}/\gamma^{*}$ at STAR complement the results of SIDIS at the HERMES \cite{HERMES:2012uyd} and COMPASS \cite{COMPASS:2013bfs,COMPASS:2017mvk} experiments and DY/$Z^{0}$ production at the E288 \cite{Ito:1980ev}, E605 \cite{Moreno:1990sf}, E772 \cite{E772:1994cpf}, CDF \cite{CDF:1999bpw,CDF:2005bdv,CDF:2012brb,CDF:2022hxs}, D0 \cite{D0:1999jba,D0:2007lmg,D0:2010dbl}, ATLAS \cite{ATLAS:2014alx,ATLAS:2015iiu,ATLAS:2019zci}, CMS \cite{CMS:2011wyd,CMS:2016mwa,CMS:2019raw}, LHCb \cite{LHCb:2015mad,LHCb:2015okr,LHCb:2016fbk}, COMPASS \cite{COMPASS:2017jbv}, and PHENIX \cite{PHENIX:2018dwt} experiments. 
On the other hand, studying the $p+p \rightarrow Z^{0}/\gamma^{*}$ process at the intermediate energies available at RHIC provides access to a higher $x$ region compared to the high energy collisions from the Tevatron and the LHC \cite{ATLAS:2010frz,ATLAS:2016fij,CMS:2010svw}. 

In addition to the unpolarized measurements, RHIC opens a window to explore the polarized TMDs through transversely polarized $p$$+$$p$ collisions.
Of particular interest is the Sivers function ($f^{\perp}_{1T}$) \cite{Sivers:1989cc,Sivers:1990fh}, which describes the unpolarized parton distribution inside a transversely polarized proton.
High precision experimental data are needed to determine $f^{\perp}_{1T}$ as current results extracted by different groups still show fairly large uncertainties for $f^{\perp}_{1T}$ \cite{Bacchetta:2011gx,Boglione:2018dqd,Cammarota:2020qcw}, especially in the relatively high $x$ region ($x\geq$ 0.1) probed by RHIC data.
There are non-trivial predictions for the process dependence of the Sivers function stemming from gauge invariance. 
In SIDIS, the Sivers function is associated with a final-state effect through gluon exchange between the struck parton and the target nucleon remnants. In $p$$+$$p$ collisions, however, the Sivers asymmetry originates from the initial state of the interaction for the DY process and $W^{\pm}/Z^{0}$ boson production. 
As a consequence, the gauge invariant definition of the Sivers function predicts the opposite sign for the Sivers function in SIDIS compared to processes with color charges in the initial state and a colorless final state, such as $p+p \rightarrow \mathrm{DY}/W^{\pm}/Z^{0}$:

\begin{equation}
f^\mathrm{SIDIS}_{q/h^\uparrow} (x, k_{\mathrm{T}}, Q^{2}) = - f^{p+p \rightarrow \mathrm{DY}/W^{\pm}/Z^{0}}_{q/h^\uparrow} (x, k_{\mathrm{T}}, Q^{2}).
\end{equation}
This non-universality of the Sivers function is a fundamental prediction from the gauge invariance of QCD and is based on the QCD factorization formalism~\cite{Collins:2002kn,Brodsky:2002rv,Ji:2002aa}.   
The experimental verification of this sign change hypothesis is a crucial measurement in hadronic physics and provides an important test of QCD factorization. 

In transversely polarized $p$$+$$p$ collisions, the Sivers function can be accessed through the transverse single spin asymmetry (TSSA) measurements in DY or $W^{\pm}$/$Z^{0}$ boson production. This asymmetry is generated from the correlation between the proton spin and the intrinsic $k_{\mathrm{T}}$ of a parton inside the proton. The amplitude of the TSSA ($A_{N}$) can be extracted from the $\phi$ modulation of ($\sigma_{\uparrow}-\sigma_{\downarrow}$)/($\sigma_{\uparrow}+\sigma_{\downarrow}$), where $\phi$ is the azimuthal angle of the measured particle and $\sigma_{\uparrow(\downarrow)}$ is its cross section with the spin direction of the proton oriented up (down) relative to the direction of its momentum. 

In this Letter, we report the first measurement of the $Z^{0}/\gamma^{*}$ differential cross section as a function of its \pT in $p$$+$$p$ collisions at a center-of-mass energy of 510\footnote[1]{The cross section measurement was performed by the STAR experiment during the 2011, 2012, 2013, and 2017 $p$$+$$p$ running periods at $\sqrt{s}$ = 500 GeV (2011 data set) and 510 GeV (2012, 2013, and 2017 data sets). The center-of-mass energy correction of 2011 data set is estimated to be 0.2\% for the combined data sets in cross section measurements, which has been ignored in this Letter.} GeV by the STAR experiment. The measurement of the $Z^{0}/\gamma^{*}$ total cross section is improved by adding a new data set compared with the previous result \cite{STAR:2020vuq}. We also report the measurement of $Z^{0}/\gamma^{*}$ $A_{N}$ in transversely polarized $p$$+$$p$ collisions at 510 GeV. These measurements are derived from studies of the $Z^{0}/\gamma^{*}\rightarrow e^{+}e^{-}$ decay channel for outgoing leptons at mid-rapidity (pseudorapidity $|\eta|<1$).

\section{Experiment and dataset}\label{sec:cuts}
The STAR detector \cite{STAR:2002eio} comprises many separate subsystems, each with specific capabilities. An essential subsystem for this measurement is the time projection chamber (TPC) \cite{Anderson:2003ur}. Together with a 0.5 T solenoidal magnetic field, the TPC provides charge discrimination and precision momentum measurements over a $|\eta|<1.3$ range with full 2$\pi$ azimuthal coverage. The barrel electromagnetic calorimeter (BEMC) \cite{STAR:2002ymp} surrounding the TPC measures the energy deposited by energetic photons and electrons with $|\eta|<1$ over the full azimuth. The $Z^{0}$ candidate events were recorded using a calorimeter trigger system which requires 12 GeV of transverse energy ($E_{T}$) in a $\Delta\eta \times \Delta\phi$ region of $\sim$ 0.1 $\times$ 0.1 of the BEMC. Primary vertices were reconstructed along the beam axis within 100 cm from the center of the STAR interaction region.

In this analysis, the differential cross section results\footnote[2]{These cross section results were obtained by averaging appropriately over the beam polarizations.} combined data samples collected in 2011, 2012, 2013, and 2017 with an integrated luminosity of 680~pb$^{-1}$. The $A_{N}$ result was measured from the data sample collected in 2017 with transversely polarized proton beams. The integrated luminosity was 340~pb$^{-1}$, which is 14 times higher than the previously published results of $A_{N}$ based on 2011 data \cite{STAR:2015vmv}. The beam polarization was determined using Coulomb-nuclear interference proton-carbon polarimeters, calibrated with a polarized hydrogen gas-jet target \cite{Jinnouchi:2004up}. The average beam polarization $\langle P \rangle$ for 2017 data was 56\%, with a relative scale uncertainty of $\Delta P/P$ = 1.4\%. 

\section{Analysis and results}

\begin{figure}[htbp]
\begin{minipage}{1.0\linewidth}
\centerline{\includegraphics[width=0.6\linewidth]{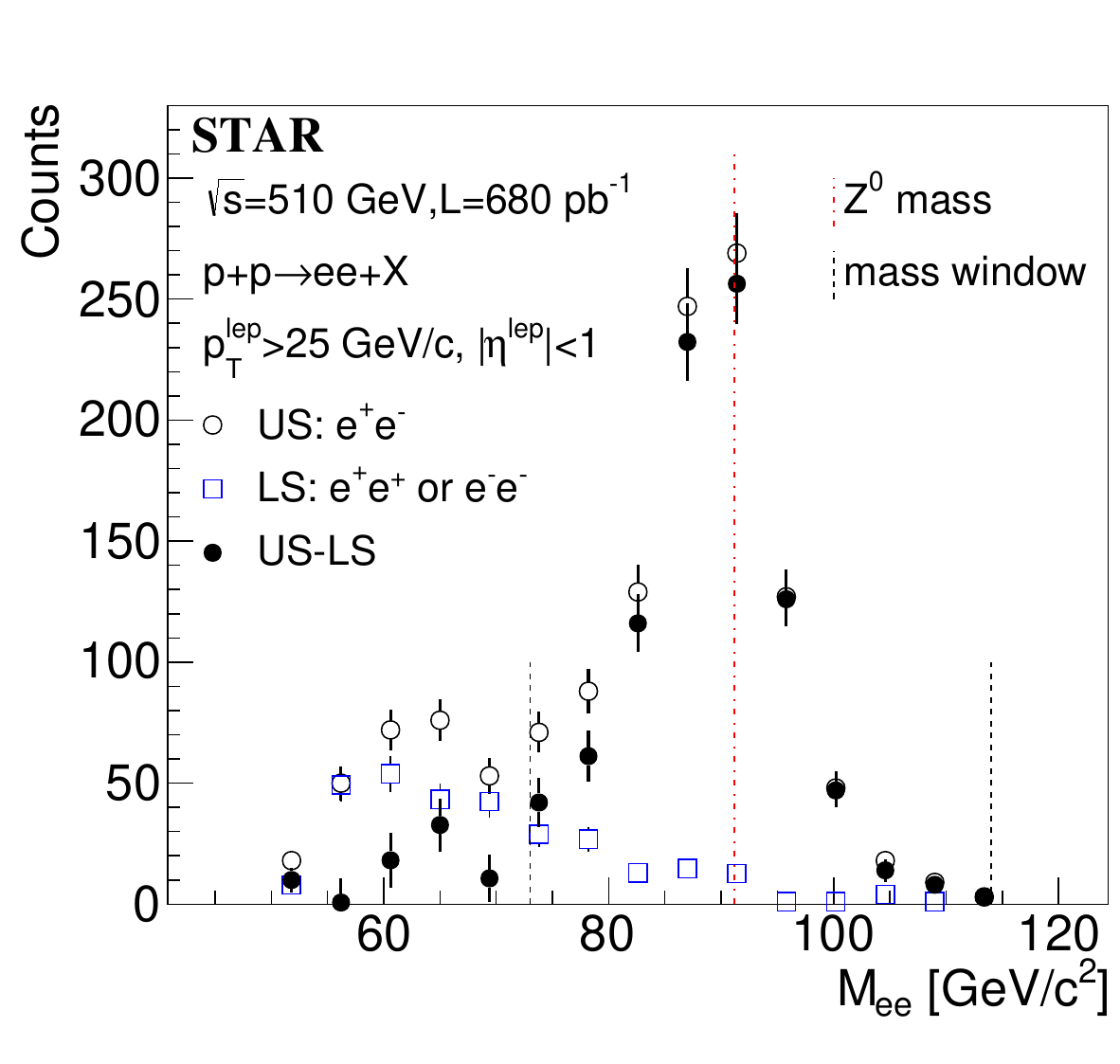}}
\end{minipage}
\caption[]{The invariant mass distribution of the reconstructed lepton pairs 
at STAR.
The open circles represent $e^{+}e^{-}$ pairs by requiring the charges of two lepton candidates to have opposite signs. The open squares represent the like-sign pairs of $e^{+}e^{+}$ and $e^{-}e^{-}$.
The solid circles represent the mass distribution after the combinatorial background subtraction.
The vertical bars indicate the statistical uncertainties. 
}
\label{fig:mass}
\end{figure}

Following exactly the same methods used in the previous measurements of $Z^{0}$ $A_{N}$ \cite{STAR:2015vmv}, $Z^{0}/\gamma^{*}\rightarrow e^{+}e^{-}$ events were selected by requiring a pair of $e^{\pm}$ candidates with opposite charge sign, $|\eta^{lep}|<1$, and $p_{\mathrm{T}}^{lep}>25$ GeV/$c$. In this analysis, we improved the measurement of the momentum of the electron and positron through scaling the angle measured by the TPC with its energy measured by the BEMC, instead of obtaining the momentum directly from the TPC. The invariant mass distribution of the $e^{+}e^{-}$ pairs is shown in Fig. \ref{fig:mass}. A signal is observed near the invariant mass of the $Z^{0}$ at $\sim$ 91 GeV/$c^{2}$. Background events, largely combinatorial in nature (uncorrelated $e^{\pm}$ pairs), were studied by requiring a pair of $e^{\pm}$ candidates with the same charge sign as shown in Fig. \ref{fig:mass} with the open squares. The solid circles represent the mass distribution after combinatorial background subtraction.

The $Z^{0}$ candidates from $e^{+}e^{-}$ were selected with a mass window cut of~73 $< M_{e^{+}e^{-}} <$ 114 GeV/$c^{2}$, the same cut as the earlier measurements \cite{STAR:2015vmv}. The candidate's transverse momentum $p^{Z^{0}}_{\mathrm{T}}$ was the vector sum of $p_{\mathrm{T}}^{lep}$ from the two decay leptons. The $p^{Z^{0}}_{\mathrm{T}}$ distribution was corrected for three effects: combinatorial background contributions; \pT unfolding due to detector resolution; and the detector inefficiencies. 
The combinatorial background correction was applied by subtracting the geometric average of the \pT distribution of $e^{+}e^{+}$ and $e^{-}e^{-}$ pairs within the mass window. The uncertainty due to this correction, estimated from the statistical uncertainties of the \pT distribution of $e^{+}e^{+}$ and $e^{-}e^{-}$ pairs, was assigned as one of the systematic uncertainties to the final $p_{\mathrm{T}}^{Z^{0}}$ spectrum.

The detector effects on the $p_{\mathrm{T}}^{Z^{0}}$ distribution were corrected by unfolding and efficiency corrections. Monte-Carlo samples generated by ``Perugia 0" \cite{Skands:2010ak} tuned PYTHIA 6.4 \cite{Sjostrand:2006za} were used at the ``particle level". The detector response for these samples were simulated using GEANT 3 \cite{Brun:1994aa}, following which the simulated events were embedded into zero-bias $p$+$p$ events and recorded with no cuts applied. The resulting event was at the ``detector level". 
An iterative unfolding technique was performed using the RooUnfold
package \cite{Adye:2011gm}, with the unfolding matrix obtained from a one-to-one mapping between the particle- and the detector-level $p_{\rm{T}}^{Z^{0}}$. The unfolding method was applied to eliminate the bin migration in $p_{\rm{T}}^{Z^{0}}$ due to momentum resolution.
The efficiency correction was then applied to the unfolded $p_{\mathrm{T}}^{Z^{0}}$ distribution. The
detector efficiency, bin by bin in $p_{\mathrm{T}}^{Z^{0}}$ for each year's data, is defined as the number of reconstructed $Z^{0}$s after the cuts divided by the
number of $Z^{0}$s from the Monte-Carlo generator level. The uncertainty of the detector efficiency correction was estimated from the statistical error of the simulated samples, which was taken as another source of systematic uncertainty of the $p_{\mathrm{T}}^{Z^{0}}$ spectrum.

The differential cross section was measured in eleven $p_{\mathrm{T}}^{Z^{0}}$ bins. Besides the contributions from the combinatorial background and efficiency corrections, the bin-by-bin systematic uncertainties were also estimated by
varying the minimum \pT requirement of the decay leptons and the uncertainty on the calibration in energy measured by the BEMC.
As described earlier, the decay lepton's \pT was required to be larger than 25 GeV/$c$. To estimate the uncertainty caused by this \pT cut, we varied the selection by requiring the lepton's \pT to be larger than 24 and 26 GeV/$c$. The relative difference of $p_{\mathrm{T}}^{Z^{0}}$ distribution, from the various selection cuts to the original one, was defined as the contribution of the \pT cut to the systematic uncertainty. 
The uncertainty of the BEMC calibration indicates how well the BEMC measures the lepton's energy. We varied the BEMC energy scale by changing the calibration gain by $\pm~$5\% for 2011$-$2013 data, the same as the published paper \cite{STAR:2020vuq}, and $\pm$ 3\% for the 2017 data. The variation of the $p_{\mathrm{T}}^{Z^{0}}$ distribution due to the gain changes was taken as the systematic uncertainty caused by the BEMC calibration uncertainty.
Generally, the dominant systematic uncertainty comes from the BEMC calibration, which varies from 4\% to 22\% in different $p_{\mathrm{T}}^{Z^{0}}$ bins. The systematic uncertainty caused by varying the minimum \pT cut is smaller than or around 3\% for most of the $p_{\mathrm{T}}^{Z^{0}}$ bins: at the highest $p_{\mathrm{T}}$, it contributes 11\% and 7\% for $p_{\mathrm{T}}^{\mathrm{min}}=24$ and 26 GeV/$c$, respectively. The contributions to the systematic uncertainty from the combinatorial background subtraction and efficiency corrections are relatively small as well, which are on average around 3\% to 4\% for all the $p_{\mathrm{T}}^{Z^{0}}$ bins. Detailed systematic uncertainties from each contribution can be found in \ref{app:sys}. Note, the $p_{\mathrm{T}}$-independent uncertainties of 10\% for $Z^{0}$ tracking efficiency and 5\% for the luminosity are not included in the $p_{\mathrm{T}}^{Z^{0}}$ spectrum, but are included in the total cross section result.

\begin{figure}[htbp]
\begin{minipage}{1.0\linewidth}
\centerline{\includegraphics[width=0.7\linewidth]{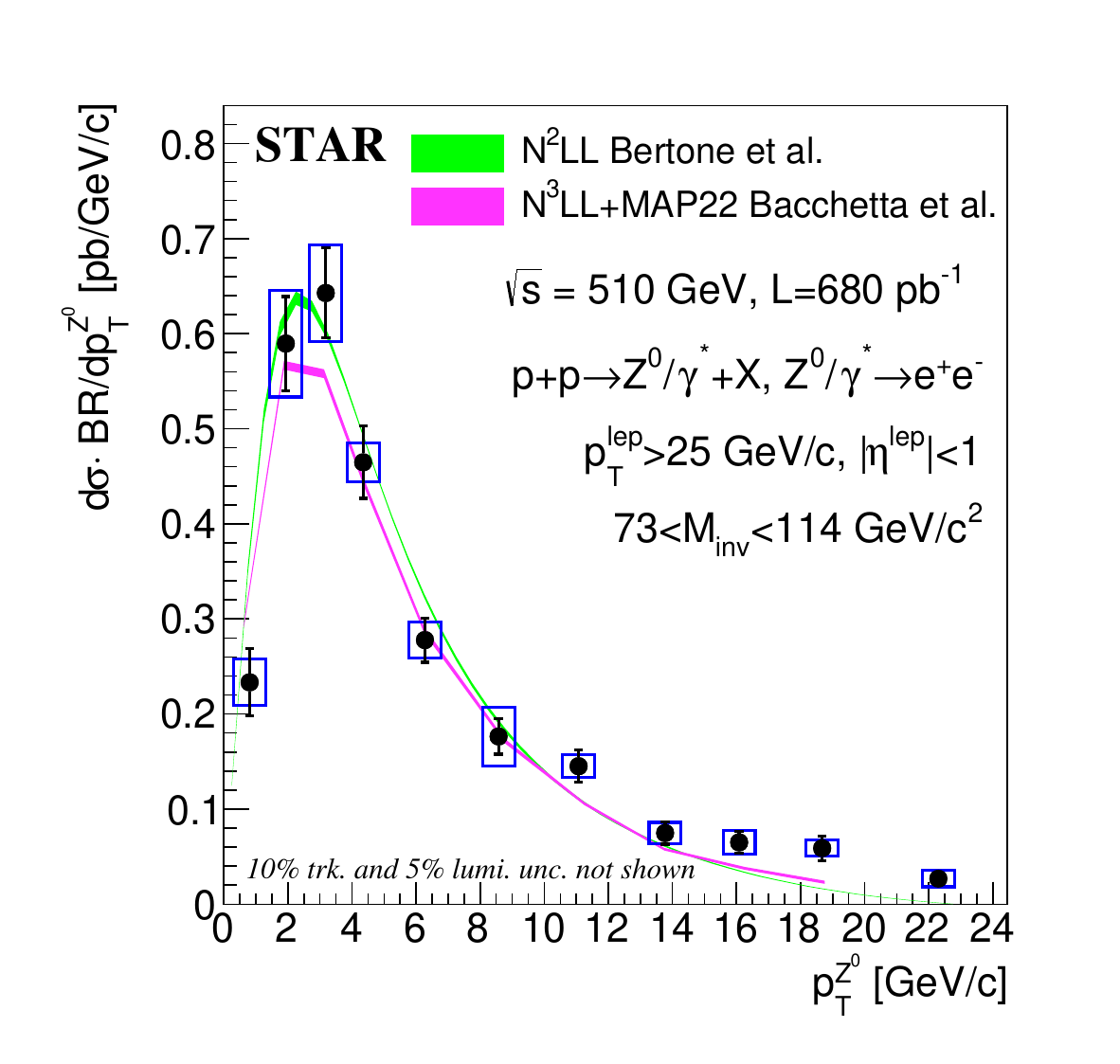}}
\end{minipage}
\caption[]{The measured $Z^{0}$ cross section as a function of its $p_{\mathrm{T}}$. 
The vertical bars indicate the statistical uncertainties and the vertical boxes indicate the systematic uncertainties. The horizontal width of the boxes is chosen for visual clarity and does not reflect the uncertainty in $p_{\mathrm{T}}^{Z^{0}}$. The $p_{\mathrm{T}}$-independent uncertainties of 10\% for $Z^{0}$ tracking efficiency and 5\% for the luminosity are not included. The result is compared with perturbative predictions at the N$^{2}$LL~\cite{Bertone:2019nxa} and N$^{3}$LL~\cite{Bacchetta:2022awv} accuracy.}
\label{fig:pT}
\end{figure}

After all the corrections and systematic uncertainty estimations described in the previous paragraphs are applied, the $Z^{0}/\gamma^{*}$ cross section as a function of its $p_{\mathrm{T}}$ is obtained and shown in Fig. \ref{fig:pT} for eleven $p_{\mathrm{T}}$ bins. BR is the branching ratio of $Z/\gamma^{*}\rightarrow e^{+}e^{-}$.
The mean value of $p_{\mathrm{T}}^{Z^{0}}$ in each bin is plotted along the horizontal axis. The plotted symbols are explained in the figure caption.
The measured $p_{\mathrm{T}}$-differential cross section of the $Z^{0}$ provides an important input to constrain the energy scale dependence of TMDs. The data are compared to calculations by two different groups: V. Bertone \textit{et al}. performed the calculation using the $\zeta -$prescription and TMD evolution at the next-to-next-to-leading order logarithmic (N$^{2}$LL) accuracy in perturbative QCD \cite{Bertone:2019nxa}; A. Bacchetta {\textit{et al.} performed the calculation using the Monte Carlo replica method and resumming large logarithms
at next-to-next-to-next-to-leading order logarithmic (N$^{3}$LL) accuracy \cite{Bacchetta:2022awv}. Data are found to be consistent with the calculations from both groups. The low $p_{\mathrm{T}}^{Z^{0}}$ spectrum is of particular relevance, since the $Q$ values should be high enough to safely apply factorization and, at the same time, $p_{\mathrm{T}}^{Z^{0}}$ should be much smaller than $Q$ in order to apply the TMD formalism. This might explain the slight discrepancy between data and the TMD-based theoretical calculations at large values of $p_{\mathrm{T}}^{Z^{0}}$. 

The $Z^{0}$ production cross sections were determined
from the sample of events which satisfy the fiducial and kinematic requirements of this analysis. 
The total fiducial cross section can be obtained by integrating the differential cross section over $p_{\mathrm{T}}^{Z^{0}}$ from Fig. \ref{fig:pT}, and is $\sigma^{\mathrm{fid}}_{Z/\gamma^{*}}\cdot \mathrm{BR} $ = 2.76 $\pm$ 0.10 (stat) $\pm$ 0.10 (sys) pb.
To determine the total production cross sections $\sigma^{\mathrm{tot}}_{Z/\gamma^{*}}$, it is
necessary to apply an acceptance correction factor, $A_{Z}$, in order to account for the fiducial and kinematic constraints imposed by the analysis. The total cross section can be written as
\begin{equation}
    \sigma^{\mathrm{tot}}_{Z/\gamma^{*}} \cdot \mathrm{BR}(Z/\gamma^{*}\rightarrow e^{+}e^{-}) = 
    \frac{\sigma^{\mathrm{fid}}_{Z/\gamma^{*}} \cdot \mathrm{BR}(Z/\gamma^{*}\rightarrow e^{+}e^{-})} { A_{Z}}.
\end{equation}\label{eq:AZ}

We applied the same method to calculate $A_{Z}$ as done in \cite{STAR:2011aa,STAR:2020vuq} based on the FEWZ program \cite{Li:2012wna}, which provides perturbative QCD calculations for $Z^{0}$ production up to order N$^{2}$L. We used the CT18 NLO PDF \cite{Hou:2019efy} as an input to obtain the value of $A_{Z}$, which is defined as the cross section ratio for the $Z^{0}$ boson with and without STAR acceptance cuts. Theoretical uncertainties in the calculation of this factor arise from several sources, including uncertainties within CT18 NLO PDF set and uncertainties on variations in the strong coupling constant, $\alpha_{s}$. The obtained $A_{Z}$ is 0.32 $\pm$ 0.01. After the kinematic acceptance correction, the total $Z^{0}$ cross section from 2011$-$2013 and 2017 data is $\sigma^{\mathrm{tot}}_{Z/\gamma^{*}}\cdot \mathrm{BR}$ = 8.63 $\pm$ 0.31 (stat) $\pm$ 0.31 (sys) $\pm$ 0.86 (eff) $\pm$ 0.43 (lumi) pb, with a relative uncertainty of 10\% for the tracking efficiency based on the past $Z^{0}$ analysis \cite{STAR:2020vuq} and 5\% for the luminosity \cite{Offlinelumi}.

\begin{figure}[htbp]
\begin{minipage}{1.0\linewidth}
\centerline{\includegraphics[width=0.7\linewidth]{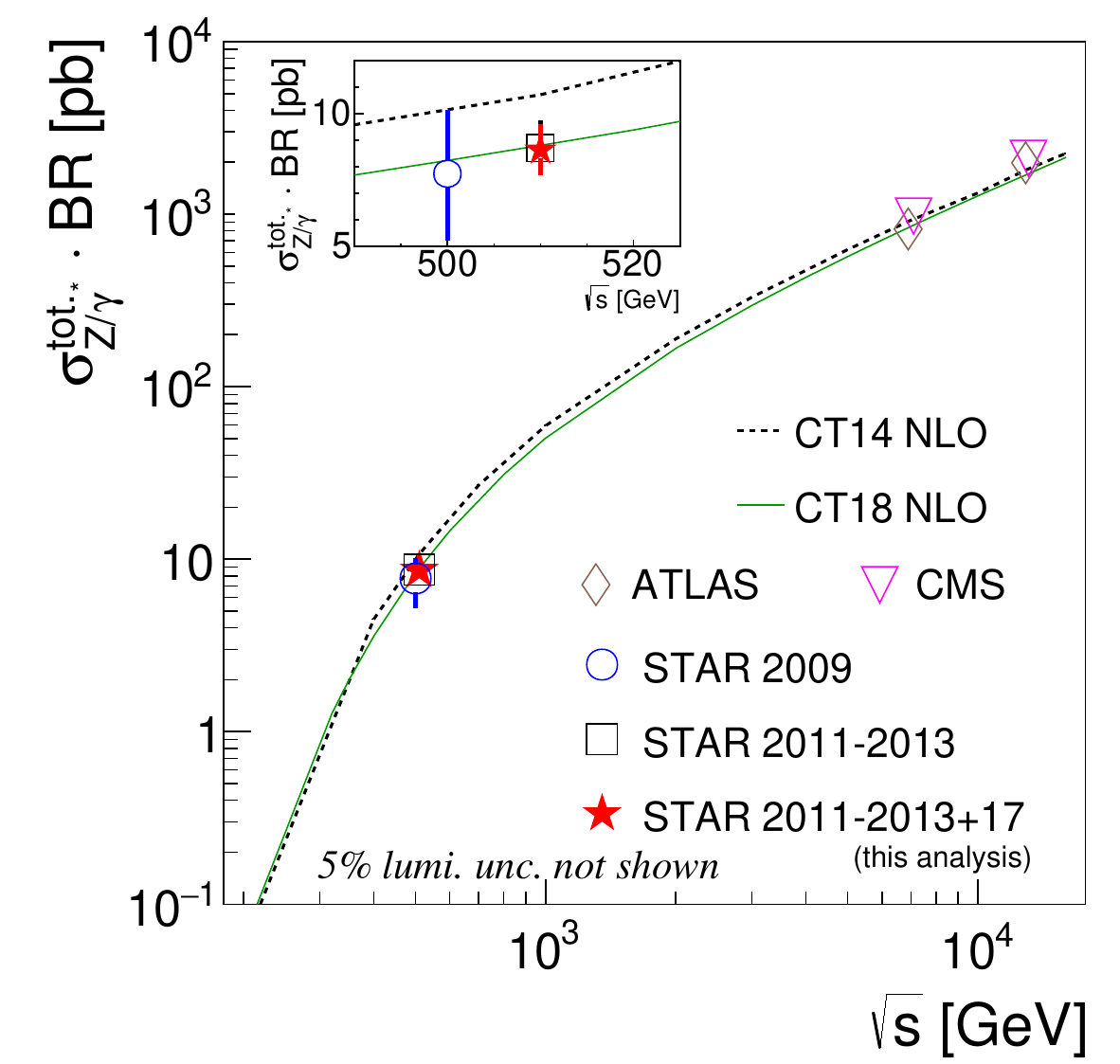}}
\end{minipage}
\caption[]{The center-of-mass energy dependence of the total $Z^{0}$ cross section compared to CT14 \cite{Dulat:2015mca} and CT18 \cite{Hou:2019efy} NLO PDF calculations. 
The measured value for the $Z^{0}$ total cross section in $\sqrt{s}=510$ GeV $p$$+$$p$ collisions is 8.63 $\pm$ 0.31 (stat) $\pm$ 0.31 (sys) $\pm$ 0.86 (eff) pb, based on a 2011$-$2013 and 2017 data sample with the integrated luminosity of 680 pb$^{-1}$. The uncertainty of 5\% for the luminosity is not included in the figure.
The previous STAR results \cite{STAR:2011aa,STAR:2020vuq} and higher energy results from the LHC \cite{CMS:2010svw,CMS:2019raw,ATLAS:2010frz,ATLAS:2016fij} are shown as well. The vertical bars indicate the total uncertainties combining statistical and systematic ones. In the small panel, the previous and current STAR results are shown within a shorter range of collision energies.}
\label{fig:xsec}
\end{figure}

Figure \ref{fig:xsec} shows the comparison of the total $Z^{0}$ cross section from this analysis with the published results from 2009 \cite{STAR:2011aa} and 2011$-$2013 \cite{STAR:2020vuq} data from STAR, higher energy $p$$+$$p$ data from the LHC \cite{CMS:2010svw,CMS:2019raw,ATLAS:2010frz,ATLAS:2016fij}, and two theoretical calculations based on CT14 and CT18 NLO PDF \cite{Dulat:2015mca,Hou:2019efy}.
The summary of the STAR results can be found in Tab. 1. 
In this analysis, 2011$-$2013 data have been reanalysed using slightly different cuts on the $Z^{0}$ mass and the lepton's $p_{\mathrm{T}}$, compared with \cite{STAR:2020vuq}.
The measured total $Z^{0}$ cross section from this analysis agrees with the previous 2009 and 2011$-$2013 results,
as shown in the small panel inside Fig. \ref{fig:xsec}. The statistical uncertainty in particular is improved significantly in this analysis compared to 2009 data. The systematic uncertainty increases in this analysis compared to the previous 2011$-$2013 result, since we considered extra contributions from the combinatorial background and efficiency corrections, 
and varying the minimum \pT requirement of the decay leptons, which were not taken in to account in \cite{STAR:2020vuq}. Additionally, a different implementation of the systematic uncertainty from the BEMC calibration was applied in this analysis. As the momentum of the decay lepton was reconstructed by scaling its energy from the BEMC, the effect of varying the BEMC gain on $p_{\mathrm{T}}$ migration is large.}
STAR data provides constraints on TMDs particularly at high $x$, since RHIC provides an intermediate collision energy. The LHC results measured at 7 and 13 TeV probe a region of $x$ lower than the STAR data at 510 GeV. Therefore, the presented STAR results are complementary to the LHC results, and provide opportunities to investigate TMD evolution as a function of $x$.
We also found all the data points to be in good agreement with the theoretical calculations.

\begin{table}[h]\label{tab:1}
\caption{Total $Z^{0}$ cross section measured from different years' data at STAR}
\centering 
\begin{tabular}{|c|c|c|} \hline
Year & Ref &
$\sigma^{tot.}_{Z/\gamma^{*}}\cdot \mathrm{BR} \pm stat_{unc.} \pm sys_{unc.} \pm {\mathrm{lumi/eff}}_{unc.}$ [pb]
\\ \hline
2009 &\cite{STAR:2011aa} & 
7.7 $\pm$ 2.1
$^{+0.5}_{-0.9}$ $\pm$ 1.0 (lumi)\\
2011-2013 & \cite{STAR:2020vuq} & 
8.7 $\pm$ 0.5 $\pm$ 0.1 $\pm$ 0.9 (eff) $\pm$ 0.8 (lumi)\\
2017 & this analysis & 
8.73 $\pm$ 0.39 $\pm$ 0.26 $\pm$ 0.87 (eff) $\pm$ 0.44 (lumi) \\
2011-2013+17 & this analysis &
8.63 $\pm$ 0.31 $\pm$ 0.31 $\pm$ 0.86 (eff) $\pm$ 0.43 (lumi)\\
\hline
\end{tabular} 
\end{table}\label{tab:1}

In addition, we report the measured $A_{N}$ of $Z^{0}$ production in $\sqrt{s}=510$ GeV $p$$+$$p$ collisions at middle rapidity ($-1<~$y$^{Z^{0}}<1$). The amplitude of the transverse single spin asymmetry of the $Z^{0}$, as described in Sec. \ref{sec:intro}, is extracted using the formula
\begin{eqnarray}
A_{N}\cdot cos(\phi) = \frac{1}{\langle P \rangle} \cdot  
\frac{\sqrt{N_\uparrow(\phi)N_\downarrow(\phi + \pi)} - \sqrt{N_\uparrow(\phi + \pi)N_\downarrow(\phi)} } 
{\sqrt{N_\uparrow(\phi)N_\downarrow(\phi + \pi)} + \sqrt{N_\uparrow(\phi + \pi)N_\downarrow(\phi)}},
\label{Eq:sqrtFormula}
\end{eqnarray}
where $N$ is the yield of $Z^0$
reconstructed
in collisions with an up/down ($\uparrow/\downarrow$) beam polarization orientation.
Defining the up transverse spin direction $\vec{S}_{\perp}$ along the $y$-axis and the direction of the incoming polarized beam $\hat{p}_{\mathrm{beam}}$ along the $z$-axis, the azimuthal angle is defined by $\vec{S}_{\perp} \cdot (\hat{p}_{\mathrm{beam}} \times \vec{p}^{Z^{0}}_{\mathrm{T}}) = |\vec{p}^{Z^{0}}_{\mathrm{T}}| \cdot \cos(\phi)$.

\begin{figure}[htbp]
\begin{minipage}{1.0\linewidth}
\centerline{\includegraphics[width=0.7\linewidth]{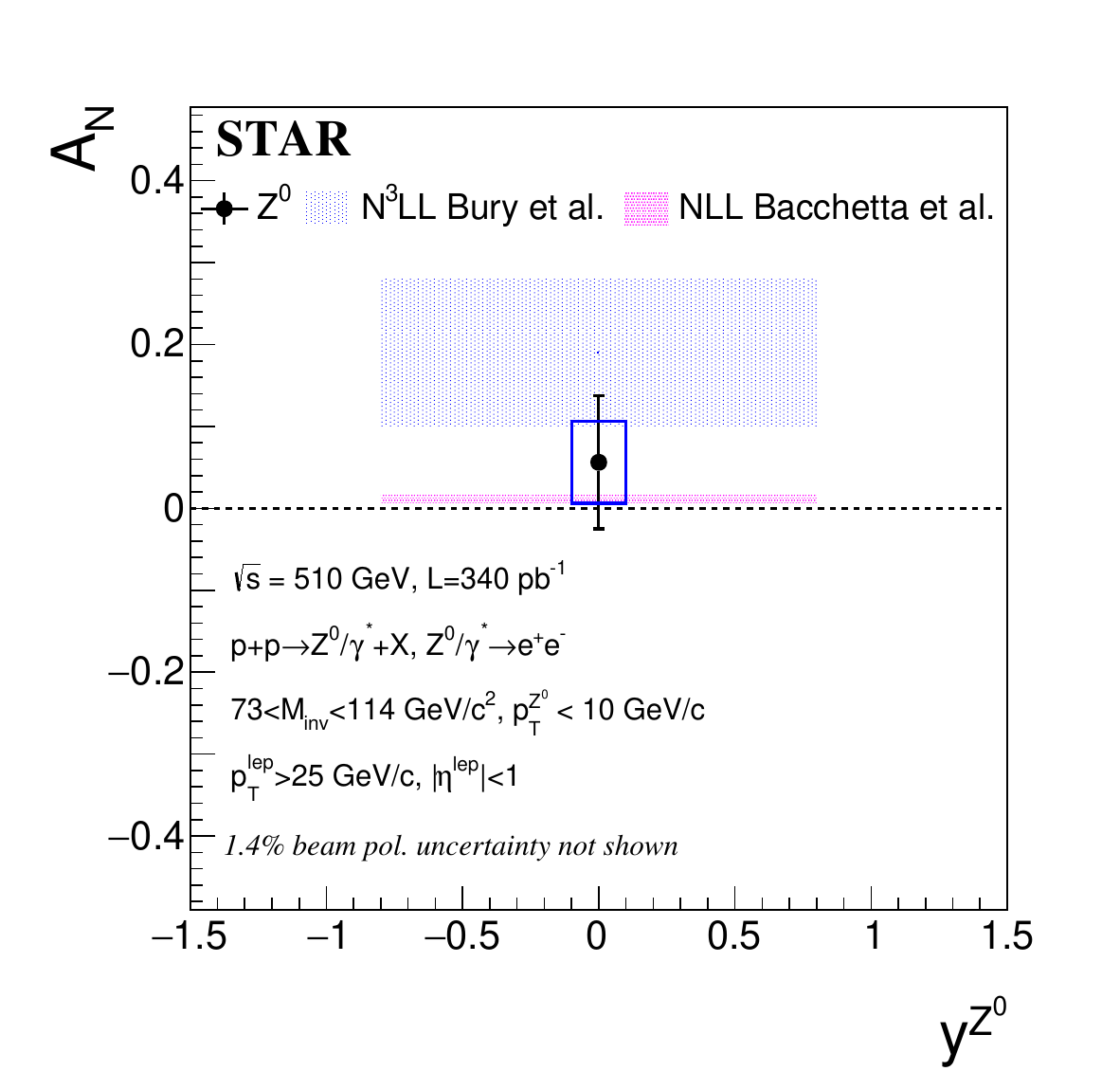}}
\end{minipage}
\caption[]{The measured $Z^{0}$ transverse single spin asymmetry in transversely polarized $p$$+$$p$ collisions, with an integrated luminosity of 340 pb$^{-1}$. 
The vertical bar indicates the statistical uncertainty and
the vertical box indicates the systematic uncertainty. The horizontal width of the box is chosen for visual
clarity and does not reflect the uncertainty in y$^{Z^{0}}$. The result is compared with two theoretical predictions, both assuming the sign change hypothesis to be true. The blue band shows the theoretical prediction calculated in the framework of TMD factorization at N$^{3}$LL accuracy \cite{Bury:2020vhj}.
The other theoretical prediction (pink band) is calculated at NLL accuracy~\cite{Bacchetta:2020gko}, in a fully consistent TMD framework.}
\label{fig:AN}
\end{figure}

The result of $A_{N}$ for the $Z^{0}$ is shown in Fig. \ref{fig:AN}. To study the TMD effects and test the sign change prediction, we limited $p_{\mathrm{T}}^{Z^{0}}$ to the range where the polarized TMD approach is applicable ($p_{\mathrm{T}}<$ 10 GeV/$c$). 
In the figure, the vertical bar indicates the statistical uncertainty and
the vertical box indicates the systematic uncertainty. The systematic uncertainty was estimated by measuring the $A_{N}$ of all like-sign pairs, which was taken as a background asymmetry. The relative uncertainty of the averaged polarization was 1.4\% and is not shown in the data point. The horizontal width of the box is chosen for visual clarity and does not reflect the uncertainty in y$^{Z^{0}}$.

This new result will provide critical input to extraction of the Sivers function, especially for valance quarks at relatively high $x$ ($x\geq$ 0.1). 
Two calculations from different groups are shown in Fig. \ref{fig:AN}, with both including the sign change hypothesis of the Sivers function. 
One is based on N$^{3}$LL accuracy of the TMD evolution in the collinear framework \cite{Bury:2020vhj}, in which, the Sivers function was expressed via an operator product expansion depending on the Qiu-Sterman function \cite{Qiu:1991pp}. The other is calculated with NLL accuracy in the traditional TMD framework \cite{Bacchetta:2020gko}, and is based on the extractions of the unpolarized and Sivers functions in a fully consistent TMD framework.
Assuming no sign change simply flips the sign of each prediction to negative, maintaining the same magnitude.
The current STAR result is not able to verify the sign change hypothesis, though it is slightly preferred over the non-sign change predictions.

\section{Summary}
We present the first measurement of the $Z^{0}$ cross section versus \pT in $p$$+$$p$ collisions at $\sqrt{s}=510$ GeV by the STAR experiment. The results combine all the data STAR has collected in 2011, 2012, 2013, and 2017, corresponding to a total luminosity of 680 pb$^{-1}$. 
The \pT spectrum of the $Z^{0}$, together with results from other experiments on DY, SIDIS, and $Z^{0}$, provide important constraints on the $x$ and $Q^{2}$ evolution as well as the process dependence of the unpolarized TMDs. A high precision measurement of the $Z^{0}$ total cross section is also reported. When combined with data from higher energy collisions, it provides a stringent test of the energy dependence of TMDs.

We also present the measurement of the $Z^{0}$ $A_{N}$ using transversely polarized $p$$+$$p$ collision data collected in 2017, corresponding to an integrated luminosity of 340 pb$^{-1}$.
The measured TSSA is 0.056 $\pm$ 0.081 (stat) $\pm$ 0.050 (sys). 
While the result can accommodate the sign change hypothesis that is based on the non-universality property of the Sivers function between DY/$Z$/$W$ production and SIDIS, it cannot conclusively verify the prediction. Precision of the $Z^{0}$ cross section and $A_{N}$ measurement will be improved using an additional 400~pb$^{-1}$ sample of \pp~data at 508 GeV that STAR collected in 2022.

\section{Acknowledgments}
We thank the RHIC Operations Group and RCF at BNL, the NERSC Center at LBNL, and the Open Science Grid consortium for providing resources and support.  This work was supported in part by the Office of Nuclear Physics within the U.S. DOE Office of Science, the U.S. National Science Foundation, National Natural Science Foundation of China, Chinese Academy of Science, the Ministry of Science and Technology of China and the Chinese Ministry of Education, the Higher Education Sprout Project by Ministry of Education at NCKU, the National Research Foundation of Korea, Czech Science Foundation and Ministry of Education, Youth and Sports of the Czech Republic, Hungarian National Research, Development and Innovation Office, New National Excellency Programme of the Hungarian Ministry of Human Capacities, Department of Atomic Energy and Department of Science and Technology of the Government of India, the National Science Centre and WUT ID-UB of Poland, the Ministry of Science, Education and Sports of the Republic of Croatia, German Bundesministerium f\"ur Bildung, Wissenschaft, Forschung and Technologie (BMBF), Helmholtz Association, Ministry of Education, Culture, Sports, Science, and Technology (MEXT) and Japan Society for the Promotion of Science (JSPS).

\newpage
\appendix
\section{Systematic uncertainties of $Z^{0}$ $p_{\mathrm{T}}$ spectrum}\label{app:sys}
\begin{table}[h]
\caption{The relative systematic uncertainties (\%) in each $p_{\mathrm{T}}$ bin from each source \label{Tab:syserrorpT} }
\centering 
\begin{tabular}{| c | c | c | c | c | c | c |} \hline
$p_{\mathrm{T}}$ bin & \makecell{Like-sign \\ correction} & \makecell{Eff.\\ correction} & \makecell{min. $p_{\mathrm{T}}^{lep}$ \\ $24~\mathrm{GeV}/c$} & \makecell{min. $p_{\mathrm{T}}^{lep}$ \\ $26~\mathrm{GeV}/c$} & BEMC gain
\\ \hline 
1 & 2.9 & 3.4 & -0.003 &  -0.04 & 13.5\\
2 &  1.0& 2.0 & 0.03 & 0.002 & 11.7\\
3 &  1.9 &1.7 & 0.14 & 0.03 & 8.0\\
4 &  1.9 & 1.7 & 0.51 & -0.08 & 4.0\\
5 &  2.2 &1.5 & 0.07 & 0.11 & 7.5\\
6 & 2.7 &1.9 & 1.2 & -1.7 & 12.8 \\
7 & 3.6 & 2.3 & 1.1 & -3.5 & 6.4 \\
8 & 6.1 & 2.7 & 0.88 & -2.8 & 4.8\\
9 & 5.1 & 3.5 & -1.1 & 0.72 & 16.4 \\
10 & 7.5 & 5.6 & 4.8 & -0.05 & 16.0 \\
11 & 7.4 & 5.4 & 11.1 & -6.8 & 22.0 \\
\hline
\end{tabular} 
\end{table}

\newpage
\bibliography{reference.bib}

\begin{thebibliography}{10}
\expandafter\ifx\csname url\endcsname\relax
  \def\url#1{\texttt{#1}}\fi
\expandafter\ifx\csname urlprefix\endcsname\relax\def\urlprefix{URL }\fi
\expandafter\ifx\csname href\endcsname\relax
  \def\href#1#2{#2} \def\path#1{#1}\fi

\bibitem{Soper:1996sn}
D.~E. Soper, {Parton distribution functions}, Nucl. Phys. B Proc. Suppl. 53
  (1997) 69--80.
\newblock \href {http://arxiv.org/abs/hep-lat/9609018}
  {\path{arXiv:hep-lat/9609018}}, \href
  {https://doi.org/10.1016/S0920-5632(96)00600-7}
  {\path{doi:10.1016/S0920-5632(96)00600-7}}.

\bibitem{Aybat:2011zv}
S.~M. Aybat, T.~C. Rogers, {TMD Parton Distribution and Fragmentation Functions
  with QCD Evolution}, Phys. Rev. D 83 (2011) 114042.
\newblock \href {http://arxiv.org/abs/1101.5057} {\path{arXiv:1101.5057}},
  \href {https://doi.org/10.1103/PhysRevD.83.114042}
  {\path{doi:10.1103/PhysRevD.83.114042}}.

\bibitem{PhysRevD.41.83}
D.~Sivers, \href{https://link.aps.org/doi/10.1103/PhysRevD.41.83}{Single-spin
  production asymmetries from the hard scattering of pointlike constituents},
  Phys. Rev. D 41 (1990) 83--90.
\newblock \href {https://doi.org/10.1103/PhysRevD.41.83}
  {\path{doi:10.1103/PhysRevD.41.83}}.
\newline\urlprefix\url{https://link.aps.org/doi/10.1103/PhysRevD.41.83}

\bibitem{Collins:1992kk}
J.~C. Collins, {Fragmentation of transversely polarized quarks probed in
  transverse momentum distributions}, Nucl. Phys. B 396 (1993) 161--182.
\newblock \href {http://arxiv.org/abs/hep-ph/9208213}
  {\path{arXiv:hep-ph/9208213}}, \href
  {https://doi.org/10.1016/0550-3213(93)90262-N}
  {\path{doi:10.1016/0550-3213(93)90262-N}}.

\bibitem{Meissner:2009ww}
S.~Meissner, A.~Metz, M.~Schlegel, {Generalized parton correlation functions
  for a spin-1/2 hadron}, JHEP 08 (2009) 056.
\newblock \href {http://arxiv.org/abs/0906.5323} {\path{arXiv:0906.5323}},
  \href {https://doi.org/10.1088/1126-6708/2009/08/056}
  {\path{doi:10.1088/1126-6708/2009/08/056}}.

\bibitem{Drell:1970wh}
S.~D. Drell, T.-M. Yan, {Massive Lepton Pair Production in Hadron-Hadron
  Collisions at High-Energies}, Phys. Rev. Lett. 25 (1970) 316--320, [Erratum:
  Phys.Rev.Lett. 25, 902 (1970)].
\newblock \href {https://doi.org/10.1103/PhysRevLett.25.316}
  {\path{doi:10.1103/PhysRevLett.25.316}}.

\bibitem{HERMES:2012uyd}
A.~Airapetian, et~al., {Multiplicities of charged pions and kaons from
  semi-inclusive deep-inelastic scattering by the proton and the deuteron},
  Phys. Rev. D 87 (2013) 074029.
\newblock \href {http://arxiv.org/abs/1212.5407} {\path{arXiv:1212.5407}},
  \href {https://doi.org/10.1103/PhysRevD.87.074029}
  {\path{doi:10.1103/PhysRevD.87.074029}}.

\bibitem{COMPASS:2013bfs}
C.~Adolph, et~al., {Hadron Transverse Momentum Distributions in Muon Deep
  Inelastic Scattering at 160 GeV/$c$}, Eur. Phys. J. C 73~(8) (2013) 2531,
  [Erratum: Eur.Phys.J.C 75, 94 (2015)].
\newblock \href {http://arxiv.org/abs/1305.7317} {\path{arXiv:1305.7317}},
  \href {https://doi.org/10.1140/epjc/s10052-013-2531-6}
  {\path{doi:10.1140/epjc/s10052-013-2531-6}}.

\bibitem{COMPASS:2017mvk}
M.~Aghasyan, et~al., {Transverse-momentum-dependent Multiplicities of Charged
  Hadrons in Muon-Deuteron Deep Inelastic Scattering}, Phys. Rev. D 97~(3)
  (2018) 032006.
\newblock \href {http://arxiv.org/abs/1709.07374} {\path{arXiv:1709.07374}},
  \href {https://doi.org/10.1103/PhysRevD.97.032006}
  {\path{doi:10.1103/PhysRevD.97.032006}}.

\bibitem{Ito:1980ev}
A.~S. Ito, et~al., {Measurement of the Continuum of Dimuons Produced in
  High-Energy Proton - Nucleus Collisions}, Phys. Rev. D 23 (1981) 604--633.
\newblock \href {https://doi.org/10.1103/PhysRevD.23.604}
  {\path{doi:10.1103/PhysRevD.23.604}}.

\bibitem{Moreno:1990sf}
G.~Moreno, et~al., {Dimuon production in proton - copper collisions at
  $\sqrt{s}$ = 38.8 GeV}, Phys. Rev. D 43 (1991) 2815--2836.
\newblock \href {https://doi.org/10.1103/PhysRevD.43.2815}
  {\path{doi:10.1103/PhysRevD.43.2815}}.

\bibitem{E772:1994cpf}
P.~L. McGaughey, et~al., {Cross-sections for the production of high mass muon
  pairs from 800-GeV proton bombardment of H-2}, Phys. Rev. D 50 (1994)
  3038--3045, [Erratum: Phys.Rev.D 60, 119903 (1999)].
\newblock \href {https://doi.org/10.1103/PhysRevD.50.3038}
  {\path{doi:10.1103/PhysRevD.50.3038}}.

\bibitem{CDF:1999bpw}
T.~Affolder, et~al., {The transverse momentum and total cross section of
  $e^+e^-$ pairs in the $Z$ boson region from $p\bar{p}$ collisions at
  $\sqrt{s} = 1.8$ TeV}, Phys. Rev. Lett. 84 (2000) 845--850.
\newblock \href {http://arxiv.org/abs/hep-ex/0001021}
  {\path{arXiv:hep-ex/0001021}}, \href
  {https://doi.org/10.1103/PhysRevLett.84.845}
  {\path{doi:10.1103/PhysRevLett.84.845}}.

\bibitem{CDF:2005bdv}
A.~Abulencia, et~al., {Measurements of inclusive W and Z cross sections in p
  anti-p collisions at $\sqrt{s}$ = 1.96 TeV}, J. Phys. G 34 (2007) 2457--2544.
\newblock \href {http://arxiv.org/abs/hep-ex/0508029}
  {\path{arXiv:hep-ex/0508029}}, \href
  {https://doi.org/10.1088/0954-3899/34/12/001}
  {\path{doi:10.1088/0954-3899/34/12/001}}.

\bibitem{CDF:2012brb}
T.~Aaltonen, et~al., {Transverse momentum cross section of $e^+e^-$ pairs in
  the $Z$-boson region from $p\bar{p}$ collisions at $\sqrt{s}=1.96$ TeV},
  Phys. Rev. D 86 (2012) 052010.
\newblock \href {http://arxiv.org/abs/1207.7138} {\path{arXiv:1207.7138}},
  \href {https://doi.org/10.1103/PhysRevD.86.052010}
  {\path{doi:10.1103/PhysRevD.86.052010}}.

\bibitem{CDF:2022hxs}
T.~Aaltonen, et~al., {High-precision measurement of the W boson mass with the
  CDF II detector}, Science 376~(6589) (2022) 170--176.
\newblock \href {https://doi.org/10.1126/science.abk1781}
  {\path{doi:10.1126/science.abk1781}}.

\bibitem{D0:1999jba}
B.~Abbott, et~al., {Measurement of the inclusive differential cross section for
  $Z$ bosons as a function of transverse momentum in $\bar{p}p$ collisions at
  $\sqrt{s} = 1.8$ TeV}, Phys. Rev. D 61 (2000) 032004.
\newblock \href {http://arxiv.org/abs/hep-ex/9907009}
  {\path{arXiv:hep-ex/9907009}}, \href
  {https://doi.org/10.1103/PhysRevD.61.032004}
  {\path{doi:10.1103/PhysRevD.61.032004}}.

\bibitem{D0:2007lmg}
V.~M. Abazov, et~al., {Measurement of the shape of the boson transverse
  momentum distribution in $p \bar{p} \to Z / \gamma^{*} \to e^+ e^- + X$
  events produced at $\sqrt{s}$=1.96 TeV}, Phys. Rev. Lett. 100 (2008) 102002.
\newblock \href {http://arxiv.org/abs/0712.0803} {\path{arXiv:0712.0803}},
  \href {https://doi.org/10.1103/PhysRevLett.100.102002}
  {\path{doi:10.1103/PhysRevLett.100.102002}}.

\bibitem{D0:2010dbl}
V.~M. Abazov, et~al., {Measurement of the Normalized $Z/\gamma^* -> \mu^+\mu^-$
  Transverse Momentum Distribution in $p\bar{p}$ Collisions at $\sqrt{s}=1.96$
  TeV}, Phys. Lett. B 693 (2010) 522--530.
\newblock \href {http://arxiv.org/abs/1006.0618} {\path{arXiv:1006.0618}},
  \href {https://doi.org/10.1016/j.physletb.2010.09.012}
  {\path{doi:10.1016/j.physletb.2010.09.012}}.

\bibitem{ATLAS:2014alx}
G.~Aad, et~al., {Measurement of the $Z/\gamma^*$ boson transverse momentum
  distribution in $pp$ collisions at $\sqrt{s}$ = 7 TeV with the ATLAS
  detector}, JHEP 09 (2014) 145.
\newblock \href {http://arxiv.org/abs/1406.3660} {\path{arXiv:1406.3660}},
  \href {https://doi.org/10.1007/JHEP09(2014)145}
  {\path{doi:10.1007/JHEP09(2014)145}}.

\bibitem{ATLAS:2015iiu}
G.~Aad, et~al., {Measurement of the transverse momentum and $\phi ^*_{\eta }$
  distributions of Drell\textendash{}Yan lepton pairs in
  proton\textendash{}proton collisions at $\sqrt{s}=8$ TeV with the ATLAS
  detector}, Eur. Phys. J. C 76~(5) (2016) 291.
\newblock \href {http://arxiv.org/abs/1512.02192} {\path{arXiv:1512.02192}},
  \href {https://doi.org/10.1140/epjc/s10052-016-4070-4}
  {\path{doi:10.1140/epjc/s10052-016-4070-4}}.

\bibitem{ATLAS:2019zci}
G.~Aad, et~al., {Measurement of the transverse momentum distribution of
  Drell\textendash{}Yan lepton pairs in proton\textendash{}proton collisions at
  $\sqrt{s}=13$ TeV with the ATLAS detector}, Eur. Phys. J. C 80~(7) (2020)
  616.
\newblock \href {http://arxiv.org/abs/1912.02844} {\path{arXiv:1912.02844}},
  \href {https://doi.org/10.1140/epjc/s10052-020-8001-z}
  {\path{doi:10.1140/epjc/s10052-020-8001-z}}.

\bibitem{CMS:2011wyd}
S.~Chatrchyan, et~al., {Measurement of the Rapidity and Transverse Momentum
  Distributions of $Z$ Bosons in $pp$ Collisions at $\sqrt{s}=7$ TeV}, Phys.
  Rev. D 85 (2012) 032002.
\newblock \href {http://arxiv.org/abs/1110.4973} {\path{arXiv:1110.4973}},
  \href {https://doi.org/10.1103/PhysRevD.85.032002}
  {\path{doi:10.1103/PhysRevD.85.032002}}.

\bibitem{CMS:2016mwa}
V.~Khachatryan, et~al., {Measurement of the transverse momentum spectra of weak
  vector bosons produced in proton-proton collisions at $ \sqrt{s}=8 $ TeV},
  JHEP 02 (2017) 096.
\newblock \href {http://arxiv.org/abs/1606.05864} {\path{arXiv:1606.05864}},
  \href {https://doi.org/10.1007/JHEP02(2017)096}
  {\path{doi:10.1007/JHEP02(2017)096}}.

\bibitem{CMS:2019raw}
A.~M. Sirunyan, et~al., {Measurements of differential Z boson production cross
  sections in proton-proton collisions at $ \sqrt{s} $ = 13 TeV}, JHEP 12
  (2019) 061.
\newblock \href {http://arxiv.org/abs/1909.04133} {\path{arXiv:1909.04133}},
  \href {https://doi.org/10.1007/JHEP12(2019)061}
  {\path{doi:10.1007/JHEP12(2019)061}}.

\bibitem{LHCb:2015mad}
R.~Aaij, et~al., {Measurement of forward W and Z boson production in $pp$
  collisions at $ \sqrt{s}=8 $ TeV}, JHEP 01 (2016) 155.
\newblock \href {http://arxiv.org/abs/1511.08039} {\path{arXiv:1511.08039}},
  \href {https://doi.org/10.1007/JHEP01(2016)155}
  {\path{doi:10.1007/JHEP01(2016)155}}.

\bibitem{LHCb:2015okr}
R.~Aaij, et~al., {Measurement of the forward $Z$ boson production cross-section
  in $pp$ collisions at $\sqrt{s}=7$ TeV}, JHEP 08 (2015) 039.
\newblock \href {http://arxiv.org/abs/1505.07024} {\path{arXiv:1505.07024}},
  \href {https://doi.org/10.1007/JHEP08(2015)039}
  {\path{doi:10.1007/JHEP08(2015)039}}.

\bibitem{LHCb:2016fbk}
R.~Aaij, et~al., {Measurement of the forward Z boson production cross-section
  in pp collisions at $\sqrt{s} = 13$ TeV}, JHEP 09 (2016) 136.
\newblock \href {http://arxiv.org/abs/1607.06495} {\path{arXiv:1607.06495}},
  \href {https://doi.org/10.1007/JHEP09(2016)136}
  {\path{doi:10.1007/JHEP09(2016)136}}.

\bibitem{COMPASS:2017jbv}
M.~Aghasyan, et~al., {First measurement of transverse-spin-dependent azimuthal
  asymmetries in the Drell-Yan process}, Phys. Rev. Lett. 119~(11) (2017)
  112002.
\newblock \href {http://arxiv.org/abs/1704.00488} {\path{arXiv:1704.00488}},
  \href {https://doi.org/10.1103/PhysRevLett.119.112002}
  {\path{doi:10.1103/PhysRevLett.119.112002}}.

\bibitem{PHENIX:2018dwt}
C.~Aidala, et~al., {Measurements of $\mu\mu$ pairs from open heavy flavor and
  Drell-Yan in $p+p$ collisions at $\sqrt{s}=200$ GeV}, Phys. Rev. D 99~(7)
  (2019) 072003.
\newblock \href {http://arxiv.org/abs/1805.02448} {\path{arXiv:1805.02448}},
  \href {https://doi.org/10.1103/PhysRevD.99.072003}
  {\path{doi:10.1103/PhysRevD.99.072003}}.

\bibitem{ATLAS:2010frz}
G.~Aad, et~al., {Measurement of the $W \to \ell\nu$ and $Z/\gamma^* \to
  \ell\ell$ Production Cross Sections in Proton-Proton Collisions at $\sqrt{s}
  = 7$ TeV with the ATLAS Detector}, JHEP 12 (2010) 060.
\newblock \href {http://arxiv.org/abs/1010.2130} {\path{arXiv:1010.2130}},
  \href {https://doi.org/10.1007/JHEP12(2010)060}
  {\path{doi:10.1007/JHEP12(2010)060}}.

\bibitem{ATLAS:2016fij}
G.~Aad, et~al., {Measurement of $W^{\pm}$ and $Z$-boson production cross
  sections in $pp$ collisions at $\sqrt{s}=13$ TeV with the ATLAS detector},
  Phys. Lett. B 759 (2016) 601--621.
\newblock \href {http://arxiv.org/abs/1603.09222} {\path{arXiv:1603.09222}},
  \href {https://doi.org/10.1016/j.physletb.2016.06.023}
  {\path{doi:10.1016/j.physletb.2016.06.023}}.

\bibitem{CMS:2010svw}
V.~Khachatryan, et~al., {Measurements of Inclusive $W$ and $Z$ Cross Sections
  in $pp$ Collisions at $\sqrt{s}=7$ TeV}, JHEP 01 (2011) 080.
\newblock \href {http://arxiv.org/abs/1012.2466} {\path{arXiv:1012.2466}},
  \href {https://doi.org/10.1007/JHEP01(2011)080}
  {\path{doi:10.1007/JHEP01(2011)080}}.

\bibitem{Sivers:1989cc}
D.~W. Sivers, {Single Spin Production Asymmetries from the Hard Scattering of
  Point-Like Constituents}, Phys. Rev. D 41 (1990) 83.
\newblock \href {https://doi.org/10.1103/PhysRevD.41.83}
  {\path{doi:10.1103/PhysRevD.41.83}}.

\bibitem{Sivers:1990fh}
D.~W. Sivers, {Hard scattering scaling laws for single spin production
  asymmetries}, Phys. Rev. D 43 (1991) 261--263.
\newblock \href {https://doi.org/10.1103/PhysRevD.43.261}
  {\path{doi:10.1103/PhysRevD.43.261}}.

\bibitem{Bacchetta:2011gx}
A.~Bacchetta, M.~Radici, {Constraining quark angular momentum through
  semi-inclusive measurements}, Phys. Rev. Lett. 107 (2011) 212001.
\newblock \href {http://arxiv.org/abs/1107.5755} {\path{arXiv:1107.5755}},
  \href {https://doi.org/10.1103/PhysRevLett.107.212001}
  {\path{doi:10.1103/PhysRevLett.107.212001}}.

\bibitem{Boglione:2018dqd}
M.~Boglione, U.~D'Alesio, C.~Flore, J.~O. Gonzalez-Hernandez, {Assessing
  signals of TMD physics in SIDIS azimuthal asymmetries and in the extraction
  of the Sivers function}, JHEP 07 (2018) 148.
\newblock \href {http://arxiv.org/abs/1806.10645} {\path{arXiv:1806.10645}},
  \href {https://doi.org/10.1007/JHEP07(2018)148}
  {\path{doi:10.1007/JHEP07(2018)148}}.

\bibitem{Cammarota:2020qcw}
J.~Cammarota, L.~Gamberg, Z.-B. Kang, J.~A. Miller, D.~Pitonyak, A.~Prokudin,
  T.~C. Rogers, N.~Sato, {Origin of single transverse-spin asymmetries in
  high-energy collisions}, Phys. Rev. D 102~(5) (2020) 054002.
\newblock \href {http://arxiv.org/abs/2002.08384} {\path{arXiv:2002.08384}},
  \href {https://doi.org/10.1103/PhysRevD.102.054002}
  {\path{doi:10.1103/PhysRevD.102.054002}}.

\bibitem{Collins:2002kn}
J.~C. Collins, {Leading twist single transverse-spin asymmetries: Drell-Yan and
  deep inelastic scattering}, Phys. Lett. B 536 (2002) 43--48.
\newblock \href {http://arxiv.org/abs/hep-ph/0204004}
  {\path{arXiv:hep-ph/0204004}}, \href
  {https://doi.org/10.1016/S0370-2693(02)01819-1}
  {\path{doi:10.1016/S0370-2693(02)01819-1}}.

\bibitem{Brodsky:2002rv}
S.~J. Brodsky, D.~S. Hwang, I.~Schmidt, {Initial state interactions and single
  spin asymmetries in Drell-Yan processes}, Nucl. Phys. B 642 (2002) 344--356.
\newblock \href {http://arxiv.org/abs/hep-ph/0206259}
  {\path{arXiv:hep-ph/0206259}}, \href
  {https://doi.org/10.1016/S0550-3213(02)00617-X}
  {\path{doi:10.1016/S0550-3213(02)00617-X}}.

\bibitem{Ji:2002aa}
X.-d. Ji, F.~Yuan, {Parton distributions in light cone gauge: Where are the
  final state interactions?}, Phys. Lett. B 543 (2002) 66--72.
\newblock \href {http://arxiv.org/abs/hep-ph/0206057}
  {\path{arXiv:hep-ph/0206057}}, \href
  {https://doi.org/10.1016/S0370-2693(02)02384-5}
  {\path{doi:10.1016/S0370-2693(02)02384-5}}.

\bibitem{STAR:2020vuq}
J.~Adam, et~al., {Measurements of $W$ and $Z/\gamma^*$ cross sections and their
  ratios in p+p collisions at RHIC}, Phys. Rev. D 103~(1) (2021) 012001.
\newblock \href {http://arxiv.org/abs/2011.04708} {\path{arXiv:2011.04708}},
  \href {https://doi.org/10.1103/PhysRevD.103.012001}
  {\path{doi:10.1103/PhysRevD.103.012001}}.

\bibitem{STAR:2002eio}
K.~H. Ackermann, et~al., {STAR detector overview}, Nucl. Instrum. Meth. A 499
  (2003) 624--632.
\newblock \href {https://doi.org/10.1016/S0168-9002(02)01960-5}
  {\path{doi:10.1016/S0168-9002(02)01960-5}}.

\bibitem{Anderson:2003ur}
M.~Anderson, et~al., {The Star time projection chamber: A Unique tool for
  studying high multiplicity events at RHIC}, Nucl. Instrum. Meth. A 499 (2003)
  659--678.
\newblock \href {http://arxiv.org/abs/nucl-ex/0301015}
  {\path{arXiv:nucl-ex/0301015}}, \href
  {https://doi.org/10.1016/S0168-9002(02)01964-2}
  {\path{doi:10.1016/S0168-9002(02)01964-2}}.

\bibitem{STAR:2002ymp}
M.~Beddo, et~al., {The STAR barrel electromagnetic calorimeter}, Nucl. Instrum.
  Meth. A 499 (2003) 725--739.
\newblock \href {https://doi.org/10.1016/S0168-9002(02)01970-8}
  {\path{doi:10.1016/S0168-9002(02)01970-8}}.

\bibitem{STAR:2015vmv}
L.~Adamczyk, et~al., {Measurement of the transverse single-spin asymmetry in
  $p^\uparrow+p \to W^{\pm}/Z^0$ at RHIC}, Phys. Rev. Lett. 116~(13) (2016)
  132301.
\newblock \href {http://arxiv.org/abs/1511.06003} {\path{arXiv:1511.06003}},
  \href {https://doi.org/10.1103/PhysRevLett.116.132301}
  {\path{doi:10.1103/PhysRevLett.116.132301}}.

\bibitem{Jinnouchi:2004up}
O.~Jinnouchi, et~al., {Measurement of the analyzing power of proton-carbon
  elastic scattering in the CNI region at RHIC}, in: {16th International Spin
  Physics Symposium (SPIN 2004)}, World Scientific, Singapore, 2004, pp.
  515--518.
\newblock \href {https://doi.org/10.1142/9789812701909\_0102}
  {\path{doi:10.1142/9789812701909\_0102}}.

\bibitem{Skands:2010ak}
P.~Z. Skands, {Tuning Monte Carlo Generators: The Perugia Tunes}, Phys. Rev. D
  82 (2010) 074018.
\newblock \href {http://arxiv.org/abs/1005.3457} {\path{arXiv:1005.3457}},
  \href {https://doi.org/10.1103/PhysRevD.82.074018}
  {\path{doi:10.1103/PhysRevD.82.074018}}.

\bibitem{Sjostrand:2006za}
T.~Sjostrand, S.~Mrenna, P.~Z. Skands, {PYTHIA 6.4 Physics and Manual}, JHEP 05
  (2006) 026.
\newblock \href {http://arxiv.org/abs/hep-ph/0603175}
  {\path{arXiv:hep-ph/0603175}}, \href
  {https://doi.org/10.1088/1126-6708/2006/05/026}
  {\path{doi:10.1088/1126-6708/2006/05/026}}.

\bibitem{Brun:1994aa}
R.~Brun, F.~Bruyant, F.~Carminati, S.~Giani, M.~Maire, A.~McPherson,
  G.~Patrick, L.~Urban, {GEANT Detector Description and Simulation Tool} (10
  1994).
\newblock \href {https://doi.org/10.17181/CERN.MUHF.DMJ1}
  {\path{doi:10.17181/CERN.MUHF.DMJ1}}.

\bibitem{Adye:2011gm}
T.~Adye, {Unfolding algorithms and tests using RooUnfold}, in: {PHYSTAT 2011},
  CERN, Geneva, 2011, pp. 313--318.
\newblock \href {http://arxiv.org/abs/1105.1160} {\path{arXiv:1105.1160}},
  \href {https://doi.org/10.5170/CERN-2011-006.313}
  {\path{doi:10.5170/CERN-2011-006.313}}.

\bibitem{Bertone:2019nxa}
V.~Bertone, I.~Scimemi, A.~Vladimirov, {Extraction of unpolarized quark
  transverse momentum dependent parton distributions from Drell-Yan/Z-boson
  production}, JHEP 06 (2019) 028.
\newblock \href {http://arxiv.org/abs/1902.08474} {\path{arXiv:1902.08474}},
  \href {https://doi.org/10.1007/JHEP06(2019)028}
  {\path{doi:10.1007/JHEP06(2019)028}}.

\bibitem{Bacchetta:2022awv}
A.~Bacchetta, V.~Bertone, C.~Bissolotti, G.~Bozzi, M.~Cerutti, F.~Piacenza,
  M.~Radici, A.~Signori, {Unpolarized transverse momentum distributions from a
  global fit of Drell-Yan and semi-inclusive deep-inelastic scattering data},
  JHEP 10 (2022) 127.
\newblock \href {http://arxiv.org/abs/2206.07598} {\path{arXiv:2206.07598}},
  \href {https://doi.org/10.1007/JHEP10(2022)127}
  {\path{doi:10.1007/JHEP10(2022)127}}.

\bibitem{Dulat:2015mca}
S.~Dulat, T.-J. Hou, J.~Gao, M.~Guzzi, J.~Huston, P.~Nadolsky, J.~Pumplin,
  C.~Schmidt, D.~Stump, C.~P. Yuan, {New parton distribution functions from a
  global analysis of quantum chromodynamics}, Phys. Rev. D 93~(3) (2016)
  033006.
\newblock \href {http://arxiv.org/abs/1506.07443} {\path{arXiv:1506.07443}},
  \href {https://doi.org/10.1103/PhysRevD.93.033006}
  {\path{doi:10.1103/PhysRevD.93.033006}}.

\bibitem{Hou:2019efy}
T.-J. Hou, et~al., {New CTEQ global analysis of quantum chromodynamics with
  high-precision data from the LHC}, Phys. Rev. D 103~(1) (2021) 014013.
\newblock \href {http://arxiv.org/abs/1912.10053} {\path{arXiv:1912.10053}},
  \href {https://doi.org/10.1103/PhysRevD.103.014013}
  {\path{doi:10.1103/PhysRevD.103.014013}}.

\bibitem{STAR:2011aa}
L.~Adamczyk, et~al., {Measurement of the $W \to e \nu$ and $Z/\gamma^* \to
  e^+e^-$ Production Cross Sections at Mid-rapidity in Proton-Proton Collisions
  at $\sqrt{s}$ = 500 GeV}, Phys. Rev. D 85 (2012) 092010.
\newblock \href {http://arxiv.org/abs/1112.2980} {\path{arXiv:1112.2980}},
  \href {https://doi.org/10.1103/PhysRevD.85.092010}
  {\path{doi:10.1103/PhysRevD.85.092010}}.

\bibitem{Li:2012wna}
Y.~Li, F.~Petriello, {Combining QCD and electroweak corrections to dilepton
  production in FEWZ}, Phys. Rev. D 86 (2012) 094034.
\newblock \href {http://arxiv.org/abs/1208.5967} {\path{arXiv:1208.5967}},
  \href {https://doi.org/10.1103/PhysRevD.86.094034}
  {\path{doi:10.1103/PhysRevD.86.094034}}.

\bibitem{Offlinelumi}
Offline analysis of luminosity calibration for $p$+$p$ 510 GeV data collected
  by the STAR expriment (2023).

\bibitem{Bury:2020vhj}
M.~Bury, A.~Prokudin, A.~Vladimirov, {Extraction of the Sivers Function from
  SIDIS, Drell-Yan, and $W^{\pm}/Z$ Data at Next-to-Next-to-Next-to Leading
  Order}, Phys. Rev. Lett. 126~(11) (2021) 112002.
\newblock \href {http://arxiv.org/abs/2012.05135} {\path{arXiv:2012.05135}},
  \href {https://doi.org/10.1103/PhysRevLett.126.112002}
  {\path{doi:10.1103/PhysRevLett.126.112002}}.

\bibitem{Bacchetta:2020gko}
A.~Bacchetta, F.~Delcarro, C.~Pisano, M.~Radici, {The 3-dimensional
  distribution of quarks in momentum space}, Phys. Lett. B 827 (2022) 136961.
\newblock \href {http://arxiv.org/abs/2004.14278} {\path{arXiv:2004.14278}},
  \href {https://doi.org/10.1016/j.physletb.2022.136961}
  {\path{doi:10.1016/j.physletb.2022.136961}}.

\bibitem{Qiu:1991pp}
J.-w. Qiu, G.~F. Sterman, {Single transverse spin asymmetries}, Phys. Rev.
  Lett. 67 (1991) 2264--2267.
\newblock \href {https://doi.org/10.1103/PhysRevLett.67.2264}
  {\path{doi:10.1103/PhysRevLett.67.2264}}.

\end{thebibliography}

\end{document}